\documentclass[aps,prl,twocolumn,superscriptaddress]{revtex4}
\usepackage{epsfig,graphicx}
\usepackage{indentfirst}
\usepackage{amsfonts,amssymb,latexsym,amsmath,enumerate,amsthm}
\usepackage{bm}
\usepackage{color}

\newcommand{\be}{\begin{equation}}
\newcommand{\ee}{\end{equation}}
\newcommand{\bea}{\begin{eqnarray}}
\newcommand{\eea}{\end{eqnarray}}
\newcommand{\dst}{\displaystyle}
\newcommand{\fr}[2]{\frac{{\dst #1}}{{\dst #2}}}

\usepackage{array}

 \newcommand{\nn}{\nonumber}

\newcommand{\bp}{{\bf p}}
\newcommand{\br}{{\bf r}}
\newcommand{\bn}{{\bf n}}

\newcommand{\bb}{{\bf b}}
\newcommand{\bE}{{\bf e}}

 \newcommand{\bB}{{\bf B}}
 
\newcommand{\bee}{{\rm e}}
\newcommand{\bii}{{\rm i}}

\begin{document}

\title{The Schwinger Scattering of Twisted Neutrons by Nuclei}
\vspace*{15px}

\author{A.\,V.~Afanasev}
\affiliation{Department of Physics,
The George Washington University, Washington, DC 20052, USA}

\author{D.\,V.~Karlovets}
\affiliation{Tomsk State University, Lenina Ave.36, 634050 Tomsk, Russia}

\author{V.\,G.~Serbo}
\affiliation{Novosibirsk State University, RUS-630090, Novosibirsk, Russia}
\affiliation{Sobolev Institute of Mathematics, RUS-630090, Novosibirsk, Russia}
\date{\today}

  %
\begin{abstract}
Thanks to J.~Schwinger, the process of elastic scattering of neutrons by nuclei is known to depend on the interference between a nuclear amplitude and an electromagnetic one for small scattering angles, resulting in spin asymmetries of a cross section or in polarization of the scattered neutrons. While this interference depends on the neutron's {\it transverse} polarization and on {\it an imaginary part} of the nuclear amplitude, this conclusion holds only for the incident plane-wave neutrons with a definite momentum. Here we show that this scattering is altered when the twisted neutrons, recently obtained experimentally, are used instead -- that is, neutrons with an orbital angular momentum. For bulk targets, the angular distributions of the scattered neutrons get modified, while scattering of a superposition of states with the different angular momenta also reveals dependence on the longitudinal polarization. For well-localized targets, the observables develop a dependence on the neutron's {\it helicity} and on {\it a real part} of the nuclear amplitude, providing full access to its phase already in the Born approximation. We argue that the corresponding spin asymmetries are measurable at existing neutron facilities. 
Thus, scattering of the twisted neutrons by nuclei can provide means for quantum tomography of the neutron states 
and become a useful tool for hadronic studies, low-energy nuclear physics, tests of fundamental symmetries, and neutron optics.
\end{abstract}
\maketitle
{\it Introduction.} -- In 1948 J.~Schwinger predicted that in a process of elastic scattering of the non-relativistic neutrons by a nucleus with a charge $Ze$ the final neutrons become spin-polarized, 
thanks to the spin-orbit coupling \cite{Schwinger-48}. Nowadays such a scattering is well studied, both theoretically \cite{Gericke08} and experimentally \cite{Shull}, 
and the polarized cold and thermal neutrons are needed for a number of standard model test measurements (see, e.g., \cite{Rauch}).

The known properties of this scattering hold only for the plane-wave neutrons with a definite momentum. 
The rapid development of laser physics, of electron microscopy, and neutron optics recently allowed one to generate photons, electrons, and neutrons in the new quantum states \cite{Airy, Allen, Airy_beam, Mono, Airy_el, Bliokh17, UFN, exp-2015}, one of which is a so-called twisted state with a phase vortex and a corresponding orbital angular momentum projection onto a propagation direction \cite{Allen, Mono, Bliokh17, UFN}. In particular, the twisted cold neutrons with the energy of $E=11$~meV and the wavelength of $0.27$~nm were experimentally obtained at NIST~\cite{exp-2015}, 
and methods for their generation were further developed in Ref.\cite{exp-2018}.
The twisted photons and electrons were shown to interact with the atoms, bulk materials, electromagnetic fields, and so on differently from the plane-wave beams 
due to a number of spin-orbit coupling phenomena, providing new tools for fundamental studies and applications \cite{Mono, Bliokh17, UFN}. 
It is of fundamental interest for the physics at mesoscopic scales to study new effects from the twisted neutrons, provided that
ultra-cold neutrons with the wavelengths of $10-100$ nm can be produced utilizing sequences of the magnetic quadrupoles \cite{exp-2018, Pushin}.

In this Letter we study how the Schwinger scattering is changed when the neutrons are twisted and identify several effects, absent with the ordinary neutron beams. 
While the twisted neutron's wave function represents a superposition of plane waves, the resultant cross section cannot generally be represented 
as an incoherent mixture of the Schwinger cross sections. It is due to this quantum self-interference of the scattering amplitudes 
that for the spatially localized targets the observables develop dependence on the neutron's longitudinal polarization and on a real part of the nuclear amplitude (and, hence, on its phase) 
-- the effects, inaccessible in the fully incoherent regime with the delocalized plane-wave neutrons.

Importantly, we show that the scattering cross section becomes dependent on the nuclear amplitude's real part already in the Born approximation, 
in contrast to the Schwinger case; and this can be useful for studying the strong interaction at low energies,
where the perturbative quantum chromodynamics is not applicable. These features are somewhat analogous to those in scattering of the relativistic twisted electrons \cite{Ivanov_Phase, Karlovets_Phase}
and reveal themselves in the spin asymmetries that can reach the values from $10^{-6}$ to $10^{-1}$, which is detectable with the existing neutron facilities \cite{Gericke08}.


{\it Scattering off a macroscopic target.} --
Let the initial neutron first be in a plane-wave state with a momentum $\bp$ and a wave function $w\,\bee^{\bii \bp \br/\hbar}$, where the spinor $w= w^{(\lambda)}(\bn)$ with a helicity $\lambda$ is normalized as $w^\dag w=1$. The final neutron's wave function is $w'\,\bee^{\bii \bp' \br/\hbar}$. We neglect the target recoil, so that $p=p'$, and introduce the unit vectors $\bn=\bp/p$ and $\bn'=\bp'/p$ with the angles $\theta, \,\varphi$ and $\theta', \,\varphi'$. The corresponding scattering amplitude is (see, for example, Ref.~\cite{BLP}, \S 42)
 \bea
 \label{1}
&& f_{\lambda \lambda'}(\bn,\,\bn')= w_{\lambda'}^{'\dag} \left(a+\bii {\bB} \bm \sigma\right) w_\lambda,
  \, \,
\bB = \beta\, \fr{\bn \times \bn'}{(\bn - \bn')^2},\nonumber \\
 &&\beta= \fr{\mu_n Z e^2}{m_p c^2}=-Z\times\,2.94\cdot 10^{-16}\;\mbox{cm},
 \eea
were $\bm \sigma$ are the Pauli matrices describing the neutron spin $\hat {\bf s}= \frac 12\, \bm \sigma$, $\mu_n=-1.91$ and $m_p$ is the proton mass. Here,  $a$ is the nuclear amplitude while $\bii {\bB} \bm \sigma$ relates to the electromagnetic interaction of the neutron's anomalous magnetic moment with a nucleus. 
For thermal neutrons with the energies near 25 meV and an $^{197}_{79}\rm Au$ nuclear target ($a$=7.63 fm \cite{NISTdata}), the relevant  parameters are
 \be
 \varepsilon \equiv |\beta/a| \approx 0.03, \; |(\mbox{Im}\,a)/a| \approx 2\cdot 10^{-4}.
 \label{parameters-for-Au}
 \ee

The standard (Schwinger) cross section summed over the spin states of the final neutrons and for the vector $\bn$ directed along the $z$-axis ({\it i.e.}, $\bn={\bf e}_3=(0,\,0,\,1)$) is
\bea
 &&\fr{d\sigma^{\rm(st)}({\bf e}_3,\,\bn', \bm \zeta)}{d\Omega'} \, =
 |a|^2+\fr 14 \left[\beta \cot(\theta'/2) \right]^2 \nonumber \\
&& -  \beta\, \zeta_\perp\,(\mbox{Im}\, a) \,\cot(\theta'/2)
 \sin(\varphi'-\varphi_\zeta).
 \label{crsection-pl}
 \eea
The interference term depends on the transverse polarization of the initial neutron $\bm \zeta_\perp=\zeta_\perp (\cos\varphi_\zeta, \sin\varphi_\zeta,0)$, but not on the longitudinal spin polarization $\zeta_z$ or the helicity $\lambda$. For small scattering angles, $\theta'\to 0$, the second term on the r.h.s. has a Coulomb-like singularity of $(1/\theta')^2$, while the third term has a singularity $1/\theta'$.

Due to the time-reversal invariance, this single-spin correlation in (\ref{crsection-pl}) is the same for either initial or final neutron polarization,
and the spin correlation averages to zero after the integration over the final neutron's azimuthal angle $\varphi'$. 
For higher energies at the LHC and RHIC hadron colliders, a similar effect of the spin Coulomb-nuclear interference \cite{Kopeliovich74, TOTEM} provided a method to measure proton beam polarization at multi-GeV energies \cite{BNL}, presently known as "CNI Polarimetry".

Now we proceed to the twisted neutrons and use an approach developed in Ref.~\cite{SIFSS-2015} for the twisted spinor particles. We assume that the incident twisted neutrons propagate along the quantization ($z$) axis and have the well--defined values of ({\it i}) a longitudinal linear momentum $p_z$, ({\it ii}) an absolute value of a transverse momentum $|{\bm p}_\perp| \equiv \hbar\varkappa$, and ({\it iii}) a projection of a total angular momentum $J_z = m$, where $m$ is a half--integer. Such \textit{a Bessel state} has, moreover, a definite energy $E = (\hbar^2\varkappa^2 + p_z^2)/(2 m_n)$, with $m_n$ being the neutron mass, and the helicity $\lambda$. The wave function is:
\begin{equation}
      \psi_{\varkappa m p_z \lambda}({\br}) = \int{\frac{{\rm d}^2 {\bp}_\perp}{(2 \pi)^2}} \, a_{\varkappa m}({\bp}_\perp) \,\bii^\lambda
   w^{(\lambda)}(\bn) \, {\rm e}^{\bii {\bp} {\br}/\hbar} \,.
   \label{Besselwf}
\end{equation}
Clearly, the function $\psi_{\varkappa m p_z \lambda}({\br})$ can be considered as a coherent superposition of the plane waves $w^{(\lambda)}(\bn) \, {\rm e}^{\bii {\bp} {\br}/\hbar}$, weighted with the amplitude
\begin{equation}
   \label{eq_a_amplitude}
   a_{\varkappa m}({\bp}_\perp) = \bii^{-m} \, {\rm e}^{\bii m \varphi} \, {\frac{2 \pi}{p_\perp}} \, \delta\left(p_\perp - \hbar \varkappa \right) \, .
\end{equation}
The momenta of these plane--wave components, $${\bp} = \left( {\bp}_\perp, p_z \right) = \left(\hbar  \varkappa \cos\varphi, \hbar \varkappa \sin\varphi, p_z \right),$$ form a surface of a cone with an opening angle
$ \theta = \arctan (\hbar \varkappa / p_z)$. 

Let us consider the scattering on a conventional thin-foil target, which we describe as an ensemble of atoms uniformly distributed over the large (compared to the beam's width) transverse extent; we call it {\it a macroscopic target}. If the target is thin, so that one can neglect the neutrons' multiple scattering and attenuation, the scattering cross section can be obtained by the averaging over the atoms' positions in the target w.r.t. the beam axis. Such an averaged cross section represents an incoherent superposition of the standard ones (see Sec. B3 in~\cite{SIFSS-2015}),
 \be
 \fr{d\bar \sigma (\theta, \theta',\varphi', \bm \zeta)}{d\Omega'}=
 \fr{1}{\cos\theta}
 \int_0^{2\pi} \fr{d\sigma^{\rm(st)}(\bn ,\,\bn',\bm \zeta)}{d\Omega'}\;
 \fr{d\varphi}{2\pi}.
 \ee
After the integration over the incoming neutron's azimuthal angle $\varphi$, we obtain (cf. Eq.(\ref{crsection-pl}))
\bea
 \label{cr-section-averaged}
 \fr{d\bar \sigma (\theta, \theta',\varphi', \bm \zeta)}{d\Omega'} &=&
   \fr{1}{\cos\theta}\Big( |a|^2 +\beta^2 \,G(\theta, \theta') \nonumber \\  &-&\beta \,(\mbox{Im}\, a)\,\zeta_\perp\, g(\theta, \theta') \,\sin(\varphi'-\varphi_\zeta)\Big), \nonumber \\
G(\theta, \theta')&=&\fr{1}{2 |\cos\theta-\cos\theta'|} - \fr 14,\;\; \\
 g(\theta, \theta')
 &=&\left\{\begin{array}{c}
    \cot(\theta'/2)\;\;\;\;\;\mbox{at}\;\;\theta'>\theta  \nonumber \\
    -\tan(\theta'/2)\;\;\mbox{at}\;\;\theta'<\theta
      \end{array}\right. .
      \eea
This cross section is still independent of $\zeta_z$ and it coincides with Eq.\eqref{crsection-pl} in the limit $\theta\to 0$. Unlike the Schwinger cross section \eqref{crsection-pl}, Eq.(\ref{cr-section-averaged}) has an angular singularity of $1/|\theta'-\theta|$ at $\theta' \to \theta$. This shift to the non-vanishing scattering angles can be useful for experimental analysis of the small-angle scattering. 
Indeed, thanks to this property the singular region is shifted from the small angles $\theta'\to 0$, which are difficult to access experimentally, to the larger ones, $\theta'\to \theta$, 
which can be controlled by the conical angle $\theta$ of the incoming neutrons. 
%

\begin{figure}[t]
	\center
	\includegraphics[width=0.85\linewidth]{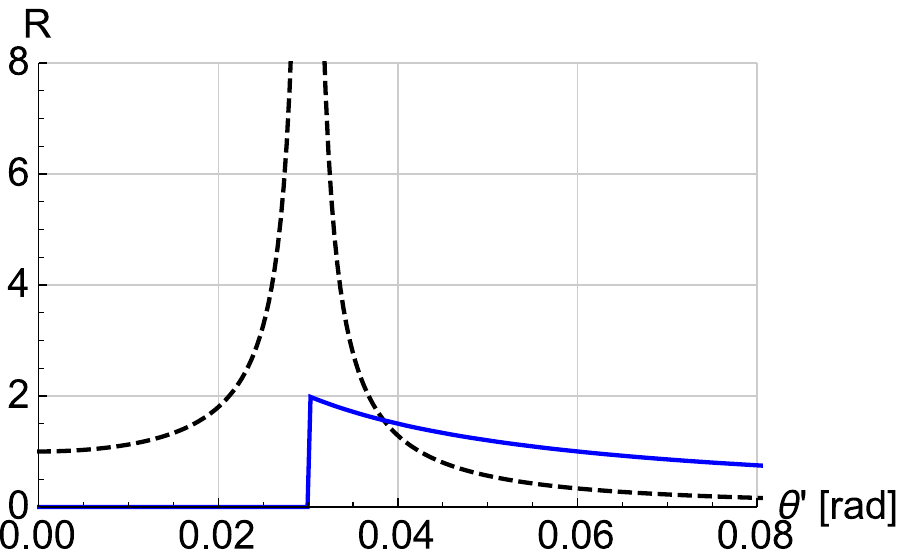}
	\caption{The functions $R_{\rm em}$ (black dashed line) and $R_{\rm int}$ (blue solid line) from
Eqs.~\eqref{Ratio} plotted vs. the neutron scattering angle $\theta'$ for $\theta=0.03$ rad 
and $\varepsilon=0.03$.
\label{Fig1}}
\end{figure}

In Fig.~\ref{Fig1} we present the functions
 \be
 R_{\rm em}=\varepsilon^2 G(\theta, \theta'),\;\;R_{\rm int}=\varepsilon g(\theta, \theta'),
 \label{Ratio}
 \ee
where $R_{\rm em}$ corresponds to a relative contribution of the electromagnetic interaction, and $R_{\rm int}$ describes interference of the electromagnetic amplitude and the nuclear one. The magnitude of the asymmetry in this case is determined by an imaginary part of the nuclear amplitude Im$(a)$, as in the original Schwinger's result, but the angular distribution is altered: the asymmetry experiences a step-like drop for the angles $\theta'\le\theta$.


{\it A macroscopic target, a superposition of two Bessel beams.} --
Let us take now a coherent superposition of two Bessel states with the different projections $m_1$ and $m_2$, but with the same helicity $\lambda$ and the same beam axis. 
Such a superposition can be generated experimentally \cite{exp-2015, exp-2018}, and it is described by the following wave function:
\bea
   \label{eq_wave_function_twisted_electrons_superposition}
   \psi^{\rm (2 \, tw)}({\br}) &=& c_1 \psi_{\varkappa m_1 p_z \lambda}({\br}) + c_2 \psi_{\varkappa m_2 p_z \lambda}({\br}),\;\; \nonumber \\
   c_n &=& |c_n| {\rm e}^{i \alpha_n} \, , \: \: \: |c_1|^2 + |c_2|^2 = 1 \, .
\eea
With the help of this expression we find
  \bea
  \label{cs-two-projection}
 &&\fr{d\bar \sigma (\theta, \theta',\varphi', \bm \zeta)}{d\Omega'} = \nn \\
&& \fr{1}{\cos\theta}\left\{ \mathcal{A}+|c_1c_2|\left(\beta^2\,\mathcal{B} +
 2(\mbox{Im}\, a)\,\beta\, (\bm\zeta{\bf C)} \right)\right\},
  \nn\\
  &&\mathcal{A} = |a|^2+\beta^2 G(\theta, \theta')+(\mbox{Im}\, a)\,\beta\, (\bm\zeta\bE_2)\,
  g(\theta, \theta'),
  \nn
  \\
  &&\mathcal{B} = \fr{\cos\gamma}{|c-c'|}\; [T(\theta, \theta')]^{|\Delta m|},
  \,
  \\
  && {\bf C} = \left[ \fr{\Delta m}{|\Delta m|}\,\left( - \fr{c'}{s'}\,\bE_1+ \bE_3\right)\,\sin\gamma
  + \fr{c-c'}{|c-c'|}\, \bE_2 \,\cos\gamma \right] \nn \\
  &&\times [T(\theta, \theta')]^{|\Delta m|},
  \nn
    \eea
where $s\equiv\sin\theta$, $c\equiv\cos\theta$, $s'\equiv\sin\theta'$, $c'\equiv\cos\theta'$ and
 \bea
 &&\gamma =(\varphi' - \pi/2)\, \Delta m +\Delta \alpha,\;\; \nn \\
 &&T(\theta, \theta')= \left(\fr{\tan(\theta/2)}{\tan(\theta'/2)}\right)^{\pm 1}
 \;\; \mbox{for}\;\; \theta' \gtrless \theta,
 \\	
&& \bE_1=(\cos\varphi',\,\sin\varphi',\, 0),\ \bE_2=(-\sin\varphi',\,\cos\varphi',\, 0),\;\; \nn \\
 &&\bE_3=(0,\,0,\, 1),\; \bE_i \bE_k=\delta_{ik}. \nn \\
 \nn
 \eea
In contrast to Eq.~(\ref{cr-section-averaged}), derived for a single-$m$ incident beam, this cross section depends on the differences of the total angular momenta, $\Delta m = m_2 - m_1\neq 0$, and of the states' phases, $\Delta \alpha = \alpha_2 - \alpha_1$. This $\Delta m$ and $\Delta \alpha$ dependence translates directly into the angular- and polarization properties of the scattered neutrons. 
In particular:

({\it i}) The cross section (\ref{cs-two-projection}) depends not only on the neutron's transverse polarization $\bm \zeta_\perp$, but also on the longitudinal one $\zeta_z$. 
It leads to the following longitudinal {\it polarization asymmetry}:
 \bea
 A_{\zeta_z}&=&\fr{d\bar \sigma (\zeta_z=+1)-d\bar \sigma (\zeta_z=-1)}
 {d\bar \sigma (\zeta_z=+1)+d\bar \sigma (\zeta_z=-1)}\nn \\ &=&
  \fr{2|c_1c_2|\,(\mbox{Im} a)\,\beta\,({\bf C} \bE_3) }{|a|^2+\beta^2 \left[
  G(\theta, \theta')+|c_1c_2|\,\mathcal{B}\right]},
  \label{A-longitudinal}
 \eea
which is of the order of $A_{\zeta_z}\sim 0.01\,\mbox{Im}\, (a)/|a|$ for the scattering angles $\theta'\approx\theta$. 
For thermal neutrons and a gold target (see Eq.~\eqref{parameters-for-Au}), the predicted asymmetry amounts to {\it a few ppm}, 
which is in a range currently accessible for experiments on the hadronic parity violation \cite{Musolf-review}, for which the above asymmetry may be a source of unwanted systematics, provided that the neutron beam becomes twisted due to uncontrolled interactions. However, as we show below, averaging over the azimuthal scattering angle $\varphi'$ eliminates the dependence on $\zeta_z$, which provides an approach to correct for this kind of systematics. Azimuthal angular coverage for the neutron-scattering detectors would be essential to deal with this systematic effect in parity-violation measurements.

\begin{figure}[t]
\centering
\includegraphics[width=0.85\linewidth]{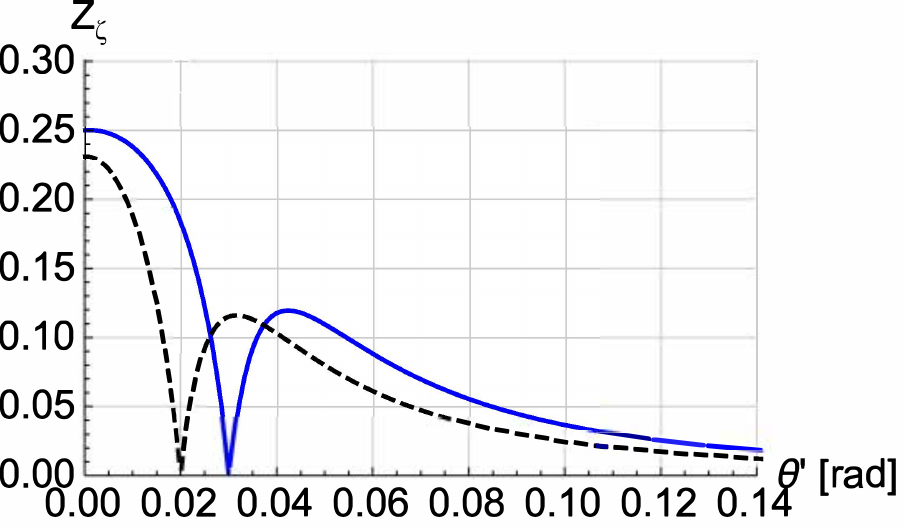}
\caption{The function $Z_\zeta$ defined from $|\langle\bm\zeta' \rangle| =\fr{|\mbox{Im}\, a|}{|a|}\;Z_\zeta$ 
(cf. Eq.~\eqref{polarization-ginal} and Eq.~\eqref{parameters-for-Au}) plotted vs. the neutron scattering angle $\theta'$
for $\varepsilon=0.03$, $2 |c_1c_2|=1$, $\Delta m=1$,  $\theta=0.03$ rad (blue solid line)
and $\theta=0.02$ rad (black dashed line).}
\label{Fig2}
\end{figure}

({\it ii}) The differential cross section~\eqref{cs-two-projection} averaged over the azimuthal angle $\varphi'$ of the final neutron depends on $\bm \zeta_\perp$ and on $\mbox{Im}\, a$ at $\Delta m =\pm 1$, but it is independent of the longitudinal polarization. It looks as follows:
 \bea
  \label{cs-two-projection-averaged}
 \left\langle
 \fr{d\bar \sigma (\theta, \theta',\varphi', \bm \zeta)}{d\Omega'}
 \right\rangle &=&
 \fr{1}{\cos\theta}\Big (|a|^2+\beta^2 G(\theta, \theta')+ \nn \\ &&
 2|c_1c_2|\, (\mbox{Im}\, a)\,\beta\, \bm\zeta \langle{\bf C}\rangle \Big),
  \eea
where
 \be
 \label{Caveraged}
  \langle{\bf C}\rangle =
     \frac 12 \, \left(\fr{c'}{s'} - \fr{c-c'}{|c-c'|} \right)
 T(\theta, \theta')\, (\cos\Delta\alpha,\, \mp \sin\Delta\alpha,\,0)
  \ee
for $\Delta m = \pm 1$, and $\langle{\bf C}\rangle =0$ otherwise. If the initial neutron is unpolarized,
then its polarization after the scattering is
 \be
 \langle\bm\zeta' \rangle = 2|c_1c_2|\, \fr{(\mbox{Im}\, a)\,\beta}{|a|^2+\beta^2 G(\theta, \theta')} 
 \, \langle{\bf C}\rangle .
 \label{polarization-ginal}
 \ee
In Fig.~\ref{Fig2} one can see that $|\langle\bm\zeta' \rangle| \sim 0.1 \; \fr{|\mbox{Im}\, a|}{|a|}$
for $\theta \approx\varepsilon$, {\it i.e.} the predicted effect amounts to {\it tens of ppm} for the thermal neutrons and the gold target (see Eq.(\ref{parameters-for-Au})).


{\it A single nucleus and a mesoscopic target.} --
Let the single-$m$ neutrons be scattered by a nucleus located in the transverse ($xy$) plane at a definite impact parameter ${\bb}= (b_x, b_y ,0)=b\,(\cos\varphi_b,\,\sin\varphi_b,\,0)$. 
Using the neutron's wave function (\ref{Besselwf}), and Eq.(\ref{1}), we find the following scattering amplitude:
 \bea
   \label{eq_scattering_amplitude_twisted_wave}
   F_{\lambda \lambda'}^{(m)}&&(\theta, \theta',\varphi', {\bb})  = \bii^{\lambda-m}
   {\rm e}^{- \bii {\bp}'_\perp {\bb}/\hbar }\, \nn \\
   &&\times\int_{0}^{2\pi} \frac{{\rm d}\varphi}{2 \pi} \,
   {\rm e}^{\bii m \varphi + \bii {\bp}_\perp {\bb}/\hbar }
   f_{\lambda \lambda'}(\bn,\,\bn'),
 \eea
where the factor ${\rm exp}(\bii {\bp}_{\perp} {\bb}/\hbar )$ specifies the lateral position of the nucleus w.r.t. the beam. 
The angular distributions of the scattered neutrons can be obtained by squaring this amplitude;
when summed up over the helicities and averaged over the azimuthal angle of the final neutrons 
they can be expressed via the quantities $\bB^{(\sigma)}(m,\varkappa, \bb)= \int_0^{2\pi} \fr{d\varphi}{2\pi}\,\bB\,
\exp\{\bii [(m-\sigma)\varphi+\varkappa b \,\cos(\varphi-\varphi_b)]\}$
and the Bessel function $J_n(z)$ as follows:
 \begin{widetext} 
\bea
 \label{AngularDistribution}
& W^{(m)}_{\lambda}(\theta, \theta',  \bb)= \sum\limits_{\lambda'} \left\langle
 \left|F_{\lambda \lambda'}^{(m)}(\theta, \theta',\varphi', {\bb}) \right|^2 \right\rangle
 = \fr 12\, \Sigma^{(m)} + \lambda\, \Delta^{(m)},
  \cr
& \Sigma^{(m)}=|a|^2 \left(J^2_{m-1/2}(\varkappa b)+ J^2_{m+1/2}(\varkappa b) \right)
  + \sum\limits_{\sigma} \left\langle (\bB^{(\sigma)*}\bB^{(\sigma)})
  -2\sigma\mbox{Im} \left(\bB^{(\sigma)*}\times \bB^{(\sigma)}\right)_z \right\rangle,
 \cr
& \Delta^{(m)}=
\left(|a|^2\cos\theta- (\mbox{Re}\, a) \,\beta\, g(\theta', \theta)\,\sin\theta \right)\,\left(J^2_{m-1/2}(\varkappa b)- J^2_{m+1/2}(\varkappa b) \right) +
 \cr
 & +  \cos\theta \sum\limits_{\sigma} \left\langle 2\sigma (\bB^{(\sigma)*}\bB^{(\sigma)})
  -\mbox{Im} \left( \bB^{(\sigma)*}\times \bB^{(\sigma)}\right)_z \right\rangle -
 \sin\theta\,\left\langle \mbox{Im} \left( \bB^{(1/2)*}\times \bB^{(-1/2)}\right)_x  -
  \mbox{Re} \left( \bB^{(1/2)*}\times \bB^{(-1/2)}\right)_y \right\rangle,
 \eea
\end{widetext}
where $\langle F \rangle  =\int_0^{2\pi} \fr{d \varphi'}{2\pi} F(\varphi')$.
As a result, we obtain a nonvanishing {\it helicity asymmetry},
 \be
 \label{Alambda}
 A_\lambda = \fr{ W_{\lambda=1/2}^{(m)} - W_{\lambda=-1/2}^{(m)} }
 { W_{\lambda=1/2}^{(m)}  + W_{\lambda=-1/2}^{(m)} } =
 \fr{\Delta^{(\rm m)}} {\Sigma^{(\rm m)}}.
 \ee

\begin{figure}[t]
\center
\includegraphics[width=0.85\linewidth]{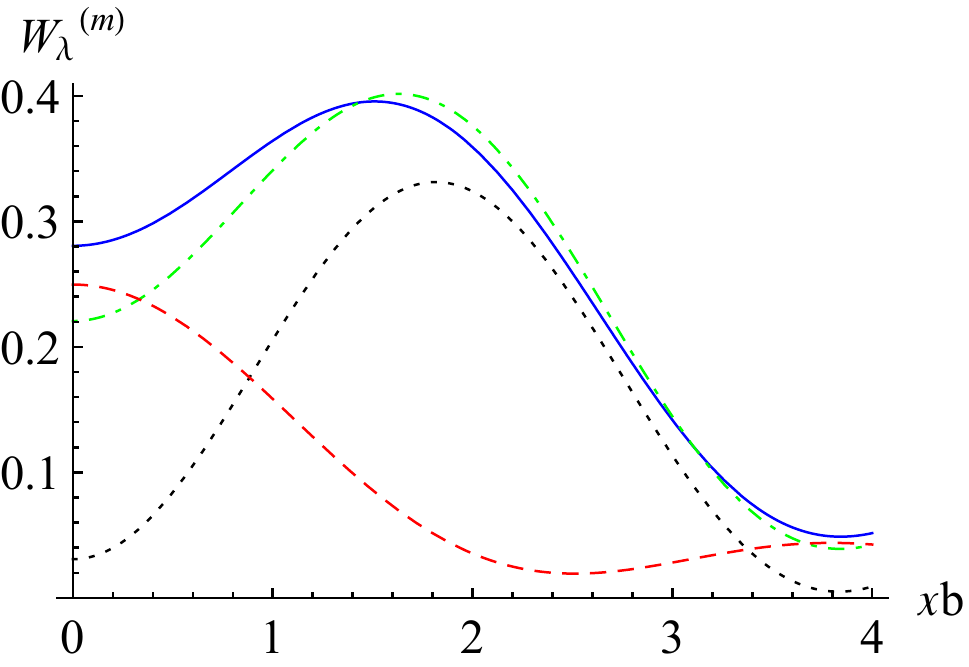}%
\caption{The probability (\ref{AngularDistribution}) as a function of the $^{197}_{79}\rm Au$ nucleus position $\varkappa b$ 
for $m = 1/2,\, \lambda = -1/2,\, \theta^{\prime} = 0.03$~rad, $\theta=0.06$~rad, and $\varepsilon=0.03$. 
The separate contributions from the nuclear amplitude (black dotted), from the electromagnetic one (dashed red), and the full result (solid blue) are shown. 
The full result with an opposite sign of the strong amplitude (green dot-dashed) demonstrates a role of its real part.}
\label{Single_Atom}
\end{figure}

\begin{figure}[t]
\center
\includegraphics[width=0.95\linewidth]{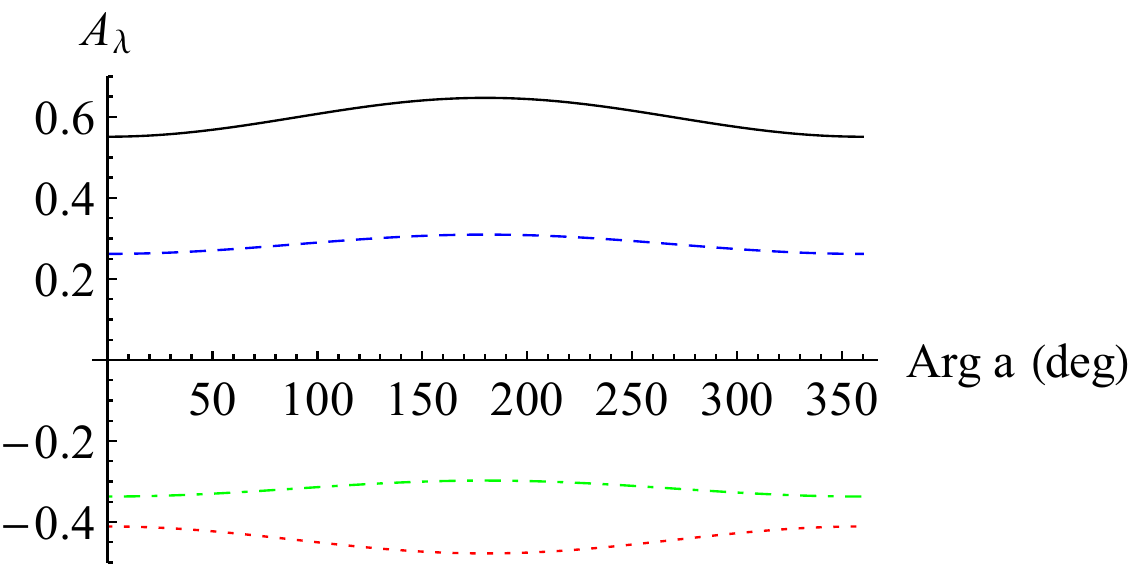}%
\caption{The helicity asymmetry (\ref{Alambda}) for $\theta^{\prime} = 0.03$~rad, $\theta=0.06$~rad, and $\varepsilon=0.03$. 
Black line: $m = 1/2, \varkappa b = 0$, blue dashed line: $m = 1/2, \varkappa b = 1$, red dotted line: $m = 1/2, \varkappa b = 2$, green dot-dashed line: $m = 5/2, \varkappa b = 1$.}
\label{Assym}
\end{figure}

In contrast to Eq.\eqref{crsection-pl}, the interference term in (\ref{AngularDistribution}) depends on the initial neutron's helicity 
and on the real part of the nuclear amplitude (and, therefore, on its phase $\text{Arg}\, a$), even after the azimuthal averaging. 
The angular distributions (\ref{AngularDistribution}) are plotted in Fig.~\ref{Single_Atom} for an $^{197}_{79}\rm Au$ nucleus as a function of its position $\varkappa b$. 
The scattering angle is chosen as $\theta^{\prime} = 0.03$~rad for which the electromagnetic and strong amplitudes
equally contribute to the cross section for the plane-wave neutrons. One can see that the former contribution 
dominates in the beam center ($b\to$0), where the interference between two amplitudes is most pronounced.
The asymmetry $A_\lambda$ is a periodic function of $\varkappa b$ and of the amplitude's phase $\text{Arg}\, a$; 
it can reach {\it tens of percent} even for a wider range of parameters than is shown in Fig.~\ref{Assym}. 
Note that outside the cone opening angle, $\theta^{\prime} > \theta$, the periodic dependence on the phase practically vanishes.

When averaging over the impact parameters ${\bf b}$, all the above features survive when the target is sub-wavelength sized or even if its width does not exceed that of the beam, $1/\varkappa$, 
much (mesoscopic), whereas for the macroscopic target they vanish, see Eq.(\ref{cr-section-averaged}).
While for a single nucleus the role of the amplitude's real part can be seen in Figs.~\ref{Single_Atom}, \ref{Assym} with the naked eye, 
the effect is additionally attenuated by the small parameter $1/(\varkappa \sigma_t) < 1$ for the mesoscopic target with the width $\sigma_t > 1/\varkappa$.
For the Gaussian target with $\sigma_t \sim 1/\varkappa - 10/\varkappa$ and the angles $\theta^{\prime} < \theta\sim 1^\circ - 10^\circ$, the asymmetry reaches the values of
\be
\label{Al}
|A_\lambda| \sim 10^{-3}-10^{-1}
\ee
for a wide range of parameters.

Thus, scattering off the well-localized targets -- say, of $\sigma_t \gtrsim 10-100$ nm in width for the neutron wave packets 
with the wavelength of $0.1 - 100$ nm and the transverse coherence length of $1/\varkappa \gtrsim 1\,\text{nm}-10\,\mu\text{m}$ \cite{exp-2018}
-- allows one to probe the nuclear amplitude's real part already in the Born approximation, 
whereas with the delocalized plane-wave neutrons such a dependence arises beyond the tree level only.
This is all the more important that the perturbative quantum chromodynamics is not applicable for these energies,
and so study of the phase effects can be useful in testing different phenomenological models.


%



{\it Conclusion. --} Scattering outcomes generally depend on the projectiles' quantum states and their spatial profiles. 
The use of the twisted neutrons in the Schwinger process makes the observables dependent on the neutron's transverse momentum 
and -- for well-localized targets -- on its orbital momentum, the helicity, and on the phase of the nuclear amplitude. 
These spin-orbit induced effects, absent with the ordinary neutrons in the Born approximation, allow one to develop tools for quantum tomography of the neutron beams 
and to study the complex nuclear amplitude as a function of the scattering angle and of the neutron energy. 
The predicted spin asymmetries range from $10^{-6}$ to $10^{-1}$ for realistic parameters 
and are detectable for existing neutron experiments aiming at much smaller numbers, up to $10^{-8}$ \cite{Gericke08}.


The helicity asymmetries for the scattering of the plane-wave neutron (or electron) beams are used to study parity violation in the nuclear force \cite{Musolf-review}. 
The asymmetries predicted here do not violate parity, as they correlate with the angular momentum, 
but the experimental methodology for studying the parity-violating scattering can also be used for the spin effects in the twisted-neutron scattering. 
Reliable estimates of these effects for specific experimental conditions require inclusion of the electron screening of the nuclear Coulomb field, similar to Ref.~\cite{Gericke08},
which will be presented elsewhere. 



We would like to thank D.~Pushin, W.\,M.~Snow, and A.~Surzhykov for useful discussions. Work of D.V.K. and V.G.S. was supported by the Russian Science Foundation (Project No.\,17-72-20013). 

%
%






\end{document}